\documentclass[preprint,usenatbib,12pt]{elsarticle}
\usepackage{amssymb}
\usepackage{multirow}
\usepackage{longtable}
\usepackage{url}
\usepackage{graphicx}	
\usepackage{amsmath}	
\usepackage{amssymb}	
\usepackage{color} 
\usepackage{footnote}
\journal{New Astronomy}

\begin{document}

\begin{frontmatter}

\title{The influence of host galaxy morphology on the properties of Type Ia supernovae from the JLA compilation}

\author[els]{V.̃~ Henne}
\ead{henne-vincent@wanadoo.fr}
\author[els,e]{M.~V.̃~Pruzhinskaya\corref{cor1}}
\ead{pruzhinskaya@gmail.com}
\author[els]{P.~Rosnet}
\ead{philippe.rosnet@clermont.in2p3.fr}
\author[els]{P.-F.~L\'{e}get}
\ead{pierre-francois.leget@clermont.in2p3.fr}
\author[els]{E.~E.~O.~Ishida}
\ead{emille.ishida@clermont.in2p3.fr}
\author[els]{A.~Ciulli}
\ead{ciulli@clermont.in2p3.fr}
\author[els]{P.~Gris}
\ead{philippe.gris@clermont.in2p3.fr}
\author[els]{L.-P.~Says}
\ead{louis-pierre.says@clermont.in2p3.fr}
\author[els]{and E.~Gangler}
\ead{emmanuel.gangler@clermont.in2p3.fr}

\cortext[cor1]{Corresponding author}

\address[els]{Laboratoire de Physique Corpusculaire, Universit\'e Clermont Auvergne, Universit\'e Blaise Pascal, CNRS/IN2P3, Clermont-Ferrand, France}
\address[e]{Lomonosov Moscow State University, Sternberg Astronomical Institute, Universitetsky pr., 13, Moscow, 119234, Russia}

\begin{abstract}
The observational cosmology with distant Type Ia supernovae (SNe) as standard candles claims that the
Universe is in accelerated expansion, caused by a large fraction of dark energy. In this paper we
investigate the SN Ia environment, studying the impact of the nature of their host galaxies on the Hubble diagram fitting. The supernovae (192 SNe) used in the analysis
were extracted from Joint-Light-curves-Analysis (JLA) compilation of high-redshift and nearby supernovae which is the best one to date.
The analysis is based on the empirical fact that SN Ia luminosities depend on their light curve shapes
and colors. We confirm that the stretch parameter of Type Ia supernovae is correlated with the host galaxy type. The supernovae with lower stretch are hosted mainly in elliptical
and lenticular galaxies. No significant correlation between SN~Ia colour and host morphology was found.

We also examine how the luminosities of SNe~Ia change depending on host galaxy morphology after stretch and colour corrections. Our results show that in old stellar populations and low dust environments,
the supernovae are slightly fainter. SNe~Ia in elliptical and lenticular galaxies have a higher $\alpha$ (slope in luminosity-stretch) and $\beta$ (slope in luminosity-colour) parameter than in spirals. However, the observed shift is at the 1-$\sigma$ uncertainty level and, therefore, can not be considered as significant.

We confirm that the supernova properties depend on their environment and that the incorporation of a host galaxy term into the Hubble diagram fit is expected to be crucial for future cosmological analyses.
\end{abstract}


\end{frontmatter}

\section{Introduction}
Type Ia supernovae (SNe) are well known as cosmological distance indicators. Among the other types of supernovae they have less dispersion at maximum light and show higher optical luminosities. Through observations of distant SNe~Ia the accelerating expansion of the Universe was discovered~\cite{1998AJ....116.1009R,1999ApJ...517..565P}. The most recent analysis of SNe~Ia indicates that considering a flat $\Lambda$CDM cosmology, our Universe is accelerating due to dark energy contamination ($\Omega_{\Lambda} = 0.705\pm0.034$~\citep{Betoule2014}). 

However, the use of SNe~Ia as ``standard candles'' would never be possible without the empirical relation between their peak luminosity and light curve shape.
This relation was independently discovered by the American statistician and astronomer B.~W.~Rust and Soviet astronomer Yu.~P.~Pskovskii in the 1970th. Using a limited sample of Type I supernovae they were able to show that the brighter the supernova, the slower its luminosity declines after achieving maximum brightness~\cite{1977SvA....21..675P,1984SvA....28..658P,1974PhDT.........7R}. Later, the dependence of colour was also reported by~\cite{1996AJ....112.2398H,1998A&A...331..815T}.  

By now several realizations of the ``standardization'' idea were developed: the $\Delta m_{15}$-method~\cite{1993ApJ...413L.105P,1999AJ....118.1766P}, the stretch-factor~\cite{1997ApJ...483..565P,1999ApJ...517..565P}, the Multicolor Light Curve Shape (MLCS)~\cite{Riess96,Jha07}, PRES~\cite{2006ApJ...647..501P}, the Spectral Adaptive Light curve Template for type Ia supernova (SALT)~\cite{Guy05,2007A&A...466...11G}, the Color-Magnitude Intercept Calibration (CMAGIC)~\cite{2003ApJ...590..944W}. Nevertheless, the SN~Ia ``standardization'' procedure is one of the main sources of systematic uncertainties in the cosmological results.

The problem is that the physical properties of SN~Ia explosions, which lead to the relation between SN luminosity and its light-curve shape and colour, are still not fully understood. It is generally believed that the SN~Ia phenomenon is an explosion of a CO white dwarf (WD) which exceeds the Chandrasekhar limit. However, the progenitor system is unclear. The accumulation of mass by a WD could be a result of either the matter accretion from a companion star (single degenerate mechanism, SD~\cite{1973ApJ...186.1007W}) or merger with another WD (double degenerate mechanism, DD~\cite{1984ApJS...54..335I,1984ApJ...277..355W}). Moreover, during the last decade the discovery of superluminous supernovae and Type Iax SNe challenges our views on possible explosion mechanisms. Without detailed spectral classification these SNe could mimic SNe~Ia.

Another important factor which could violate the ``standard candle'' hypothesis is dust. Dust around the supernovae, as well as in the host galaxy, surely affects light curve behavior. The distribution and properties of dust in host galaxies of supernovae could be different from that in the Milky Way. Furthermore, one noteworthy ambiguity is connected to the so-called gray dust: its absorption is wavelength independent and essentially cannot be taken into account. Gray dust can
be large dust particles with typical sizes greater than 0.01~$\mu$m~\cite{1999ApJ...512L..19A}. This could lead to the dimming of distant supernovae and mimic the accelerated expansion~\cite{2011ARep...55..497B}. Indeed, the amount of such dust inside host galaxies of SNe is proportional to the star formation rate, which increases into the past, and could produce an apparent fall in the luminosity of distant supernovae. However, in~\cite{2011AstL...37..663P} it was shown that SNe which are free from gray dust absorption, i.e. ``pure'' SNe, also indicate the accelerating expansion. 

At last, the metallicity of the progenitor defines the Chandrasekhar limit. A lower metallicity involves an increase of the Chandrasekhar limit and could affect the explosion process. 

The above-listed factors could be considered as environmental effects. In the epoch of large transient surveys, development of observational techniques,  and further processing of the data, a study of environment is of high priority. 

In this paper we consider the properties of SNe and their impact on Hubble diagram in different environments. We separate Type Ia supernovae according to the type of host galaxy into three subsamples: supernovae that exploded in elliptical/lenticular (E-S0), early-type spiral (Sa-Sc), and late-type spiral (Sd-Ir) galaxies. The idea of our approach is close to the idea described in~\cite{2003MNRAS.340.1057S,2011AstL...37..663P}. The reasoning behind this idea is as follows.
First, the oldest, i.e. metal-poor, stars with an age comparable to that of the Universe lie in elliptical/lenticular galaxies. 
This automatically leads to a more homogeneous chemical composition of the progenitor stars. Second, the dust (including gray dust) is almost absent in these regions. And thirdly, SNe in elliptical galaxies probably have a common explosion mechanism --- DD. Indeed, while in elliptical galaxies it is anyway
the white dwarf merger mechanism which provides up to 99\% of SN Ia explosions
(see~\cite{2011NewA...16..250L}), this is not the case for spiral galaxies. In the latter galaxies some fraction of SNe Ia can explode via the SD mechanism which suggests the
accumulation of mass by a white dwarf from an intermediate-mass star in a binary system. The considered types of the galaxies differ in star formation rate, which is low for elliptical/lenticular galaxies and high for spirals. However, while in early-type spirals star formation regions are located in spiral arms, in late-type they are not associated with any particular structures. 

Our analysis is based on the Joint Light-curve Analysis (JLA) sample of supernovae which includes 740 spectroscopically confirmed Type Ia supernovae~\cite{Betoule2014}. The JLA sample is the best SNe sample to date. The main advantages of JLA compared to previous compilations are: an intercalibration between different surveys and a thorough investigation of systematic uncertainties. The presented analysis is the first study of JLA supernovae depending on host galaxy morphology.

The paper is organized as follows. In Section 2 we describe how we make a sample of SNe~Ia for our analysis and how we divide them considering the morphology of the host galaxy. Section 3 contains the description of the Hubble-diagram fitting procedure. In Section 4 we perform the analysis of impact of the host morphology on SN~Ia properties and compare our results with those of earlier studies. The conclusions are presented in Section 5. 
 
\section{Type Ia supernovae and host galaxy data}
\subsection{JLA compilation}
The JLA compilation~\citep{Betoule2014} regroups 740 nearby and distant supernovae up to redshift $1.3$, that has been acquired between 1990 and 2008. In this sample, 374 objects come from the Sloan Digital Sky Survey (SDSS), 239 from the Supernovae Legacy Survey (SNLS), and 9 from the Hubble Space Telescope (HST). The remaining 118 are low-redshift supernovae measured by different nearby experiments. The JLA supernovae were standardized using the SALT2 model, an empirical model of spectro-photometric time evolution of SN~Ia~\citep{2007A&A...466...11G}.

The following data for each supernova were extracted from JLA to perform the analysis: 
\begin{itemize}
\item $z_{\rm cmb}$: the redshift in the CMB frame with its uncertainty;
\item $z_{\rm hel}$: the redshift in the heliocentric frame;
\item $m_{\rm B}^{*}$: the value of the $B$-band peak magnitude with its uncertainty;
\item $X_1$: SALT2 shape parameter (stretch) with its uncertainty;
\item $C$: SALT2 colour parameter with its uncertainty;
\item ra: right ascension in degrees (J2000);
\item dec: declination in degrees (J2000).
\end{itemize}

\subsection{Host galaxy morphology}
The morphological classification of galaxies proposed by Hubble~\citep{1926ApJ....64..321H,1936rene.book.....H} and after developed by de Vaucouleurs~\cite{1959HDP....53..275D}  and Sandage~\cite{1961hag..book.....S} is strongly correlated with physical properties of the galaxies such as colour, mean surface brightness, neutral hydrogen content etc (see~\cite{2013pss6.book....1B} for a review).

For our analysis it is important that the stellar population, amount of dust, star formation rate also evolve along the morphological sequence. Elliptical galaxies are dominated by old-stars and they are relatively dust-free. Spiral galaxies are characterized by the appearance of the spiral arms with star formation regions and diversity of stellar population. Irregular galaxies like spirals have old and young stellar populations but the star formation occurs in the absence of spiral arms~\cite{1997PASP..109..937H}.  

\begin{table}[h]
 \noindent\makebox[\textwidth]{
 \begin{tabular}{|c|c|c|c|c|c|c|c|c|c|c|c|c|c|c|c|c|}
 \hline
  \multicolumn{6}{|c|}{Elliptical/Lenticular (E/L)} & \multicolumn{6}{c|}{Early type spiral (ES)} & \multicolumn{5}{c|}{Late type spiral (LS)}\\
 \hline
 cE & E & $E^{+}$ & $S0^{-}$ & $S0^{0}$ & $S0^{+}$ & S0/a & Sa & Sab & Sb & Sbc & Sc & Scd & Sd & Sdm & Sm & Im \\
 \hline
 \end{tabular}}
  \caption[Morphological classifications]
 { Morphological classification of galaxies implemented in the paper.}
 \label{table:morphoclass}
 \end{table}

We divided the host galaxies of supernovae into three groups: elliptical/lenticular (E/L), early-type spiral (ES), and late-type spiral (LS) galaxies (see Table~\ref{table:morphoclass}): 

\begin{itemize}
 \item E/L: in this category elliptical and lenticular galaxies were included. Elliptical galaxies have a shape of ellipse or sphere, without apparent substructure. The surface brightness is uniform in the core and declines suddenly outwards.  Lenticular galaxies have a spheroidal bulge and a disk, without spiral arms.
 \item ES: galaxies with type from S0/a to Sc belong to the early-type spirals. This type is characterized by a prominent bulge, well formed arms and dust lanes. 
 \item LS: the late-type spirals (Scd-Im) have loosely wound and weak arms, and a weak bulge. Both, arms and bulge completely disappear in irregular galaxies (Im)~\cite{2011A&A...532A..74B,2013pss6.book....1B}.
\end{itemize}

Our choice was limited by statistics and difficulties connected with morphological classification. While  confident detailed classification is possible only for nearby galaxies, to classify the distant galaxies usually the comparison between the observed spectra and standard spectra templates for each morphological type is applied.  However, even for nearby galaxies sometimes there is a contradiction between different catalogs. For instance, the host galaxy of nearby SN~1993O in our sample is classified as S0 and Sb by different authors~\cite{2005ApJ...624..532R,2009ApJ...707.1449N}.

To classify host galaxies of the SDSS supernovae we used the data provided by~\cite{2008AJ....135.1766Z}. Zheng et al. have developed automatic classification programs to define SN and host-galaxy types as well as precise redshifts, exploring template cross-correlation and principal component analysis. The morphological types for host galaxies, independently provided by both techniques, are given in Table 3 and 4 of~\cite{2008AJ....135.1766Z}. We made a cross-correlation between JLA SNe and Zheng et al. data; 77 coincidences were found. For 18 galaxies~\cite{2008AJ....135.1766Z} provides several morphological types. In such case we  adopted the type obtained by cross-correlation method. If cross-correlation gives several possible types we chose the one corresponding to the highest best-fit criterion.

The host morphology for nearby supernovae was extracted from different sources~\cite{2005ApJ...624..532R,2009ApJ...707.1449N,2011A&A...532A..74B,2013AJ....146...86T}. If different catalogs provided the different morphological types for the same galaxy, the final decision was rendered by visual examination of images.  

We did not find in the literature the detailed morphology of hosts of SNLS SNe in the JLA sample. 

The classification of JLA HST supernovae can be found in~\cite{2005ApJ...635.1022C,2012ApJ...750....1M}. However, this classification divides hosts on early-type (E/S0) and late-type (S) galaxies that is not enough for our analysis. 

Therefore, the detailed host classification was found for 192 SN~Ia from JLA sample (see~Table~\ref{table:stat}). The full list of 192 SNe is presented in Appendix (Table~\ref{SN_list}).

\begin{table}[h]
 \noindent\makebox[\textwidth]{
 \begin{tabular}{lc}
 \hline
 Host type & Number of SNe\\
 \hline
E/L& 39 \\
ES& 88 \\
LS& 65 \\
\hline
Total: & 192\\
\hline
 \end{tabular}}
  \caption[Morphological classifications]
 {Final sample of SN~Ia relative to host galaxy type. The redshift range is $0.01<z<0.4$ for the whole sample.}
 \label{table:stat}
 \end{table}
 
\section{Hubble diagram fitting} 
The cosmological analysis is based on empirical relations between the light curve parameters and absolute magnitude of supernovae~\cite{1998A&A...331..815T}. We use SALT2 parameters $X_1$ describing the time stretching of the light-curve and the colour offset with respect to the average at the date of maximum luminosity in the $B$-band, $C = (B-V)_{\rm max}-\langle B-V \rangle$. 

In this analysis we adopt the classical standardization of the distance modulus:

\begin{equation}
\mu = m_{\rm B}^{*} - (M_{\rm B} - \alpha X_{1} + \beta C),
\label{eq:distanceluminosity}
\end{equation}
where $M_{\rm B}$ is a standardized absolute magnitude of the SN~Ia in the $B$-band for $X_1=C=0$, $\alpha$ and $\beta$ describe, consequently, the stretch and colour law for the whole sample. The standardized absolute magnitude of a SN can be defined as $M  = M_{\rm B} - \alpha X_{1} + \beta C$. The value of the $B$-band apparent magnitude $m_{\rm B}^{*}$ at maximum light, stretch factor $X_1$ and colour $C$ are taken directly from the JLA dataset~\cite{Betoule2014}, while the three nuisance parameters $M_{\rm B}$, $\alpha$ and $\beta$ are free parameters of the Hubble diagram fit, as well as the cosmological parameter $\Omega_\Lambda$.
All free parameters are determined by minimizing the following chi-square:

\begin{equation}
\chi^{2} = [\mu(M_{\rm B},\alpha,\beta)-\mu_{\rm theory}(\Omega_{\Lambda})]^{\dagger}V^{-1}[\mu(M_{\rm B},\alpha,\beta)-\mu_{\rm theory}(\Omega_{\Lambda})],
\end{equation}
where $\mu$ is given by equation~\ref{eq:distanceluminosity}, $\mu_{\rm theory} = 5\lg d_{\rm L} - 5$, where $d_{\rm L}$ is the luminosity distance in parsecs. $V$ is a full covariance matrix of the vector of the distance modulus estimates that includes uncertainties in cosmological redshift due to peculiar velocities, the magnitude variations due to gravitational lensing, the uncounted intrinsic variation in SN magnitude, and different systematic errors such as calibration and bias uncertainties (for the full description of the covariance matrix see~\cite{Betoule2014}). The Hubble constant is fixed, $H_0 = 70$ km~s$^{-1}$~Mpc$^{-1}$, and the fit assumes a flat Universe dominated by its matter content $\Omega_{\rm m}$ and the dark energy $\Omega_\Lambda$, i.e. such that $\Omega_{\rm m} + \Omega_{\Lambda} = 1$. 

The fit we applied is absolutely the same as described in~\cite{Betoule2014}. However, in the JLA fit one more additional parameter was introduced. It was assumed that $M_{\rm B}$ is related to the host stellar mass by a simple step function:

\begin{equation}
M_{\rm B} =\begin{cases} M_{\rm B} & \text{if }  M_{\rm stellar} <10^{10}M_{\odot} \\
                     M_{\rm B} + \Delta_{\rm M} &  \text{   otherwise.} 
       \end{cases}       
\label{eq:HGstepfunction}
\end{equation}

To study the environmental effects of SNe Ia we removed this correction as well as the corresponding component of the full covariance matrix from our fitting procedure. 

The $\chi^{2}$ has been minimized with the minimizer ``iminuit'' of python 2.7. 

Due to the fact that we use only a part of the JLA sample (192 SNe), first, we checked that our sample is unbiased by comparing the Hubble diagram fit of the selected sample to the full JLA dataset, as shown in Table~\ref{table:2fullfit}.
The results are compatible within one standard deviation, except for the $\beta$ parameter which varies up to about 2-$\sigma$.
This difference can express the fact that the selected sample of SNe Ia is not fully representative of the full JLA dataset in term of colour. It could be also connected to the colour evolution with redshift~\cite{2011ApJ...740...72M}.
One important point is that the selected sample gives a cosmological result in agreement with the full JLA dataset.
To examine the systematic effects due to redshift we made the plot of the redshift distribution of the three types of galaxies (see Fig.~\ref{fig:Host_mass}, left plot). For all three subsamples the redshift is distributed in the same way. 

\begin{table}[h]
 \noindent\makebox[\textwidth]{
\begin{tabular}{lcc}
  \hline
 &    Full JLA sample& Our sample\\
 \hline
Number of SNe & $\,\;\;704$ & $\,\;\;\;192$\\
$z_{\rm max}$ & $\,\;~1.3$ & $\,\;~0.4$\\
$\chi^{2}$/dof & $\,\;~\;\,0.99$ &  $\,\;~\;\,0.79$\\
$\Omega_{\Lambda}$ &  $\,\;~\;\,0.720 \pm 0.033$ & $\,\;~\;\,0.740\pm 0.088$\\
$\alpha$ & $\,\;~\;\,0.132 \pm 0.006$ & $\,\;~\;\,0.137 \pm 0.011$\\
$\beta$ & $\,\;~\;\,3.12 \pm 0.08$ & $\,\;~\;\,2.75 \pm 0.15$\\
$M_{\rm B}$ & $-19.08 \pm 0.02$ & $-19.11 \pm 0.03$\\
  \hline
 \end{tabular}}
   \caption[Result full fit]
 {Best-fit parameters for the full JLA dataset and the sample extracted from JLA and used in the current analysis. $\Delta_{\rm M} = 0$ in both cases.}
 \label{table:2fullfit}
\end{table}

\section{The study of environmental effects}
\subsection{Stretch and colour}
Before studying the impact of the host galaxy morphology on the Hubble diagram fitting we analyze the stretch and colour distributions for each SN subsample. 

The correlation between the stretch-factor and host galaxy properties was found in previous studies~\cite{1995AJ....109....1H,1996AJ....112.2398H,1999AJ....117..707R,2000AJ....120.1479H,2003MNRAS.340.1057S}. It was established that the SNe~Ia with the slowest decline rate, therefore, the most luminous ones, appear in galaxies with a younger stellar population (spiral and irregular galaxies). 

\begin{table}[h]
 \noindent\makebox[\textwidth]{
 \begin{tabular}{lcccc}
 \hline
 & $\langle X_1 \rangle$ & $X_1$ RMS & $\langle C \rangle$ & $C$ RMS \\
\hline 
Our sample& $-0.19 \pm 0.08$ & $1.07 \pm 0.06$ & $-0.009 \pm 0.006$ & $0.081 \pm 0.004$ \\
\hline
E/L& $-0.93 \pm 0.17$ & $1.06 \pm 0.12$ & $-0.023 \pm 0.010 $ & $0.060 \pm 0.007 $   \\
ES& $-0.18 \pm 0.11$ & $1.04 \pm 0.08$ & $-0.008 \pm 0.009$ & $0.088 \pm 0.007$ \\
LS& $\,\;~0.25 \pm 0.11$ & $0.86 \pm 0.08$ & $-0.003 \pm 0.010$ & $0.081 \pm 0.007$ \\
\hline
 \end{tabular}}
 \caption{{The mean values and RMS for stretch $X_1$ and colour $C$ for our full sample and the three subsamples.}}
\label{table:SALT2_Morpho}
\end{table}
 
\begin{figure}
\begin{center}
  \includegraphics[width = 15cm]{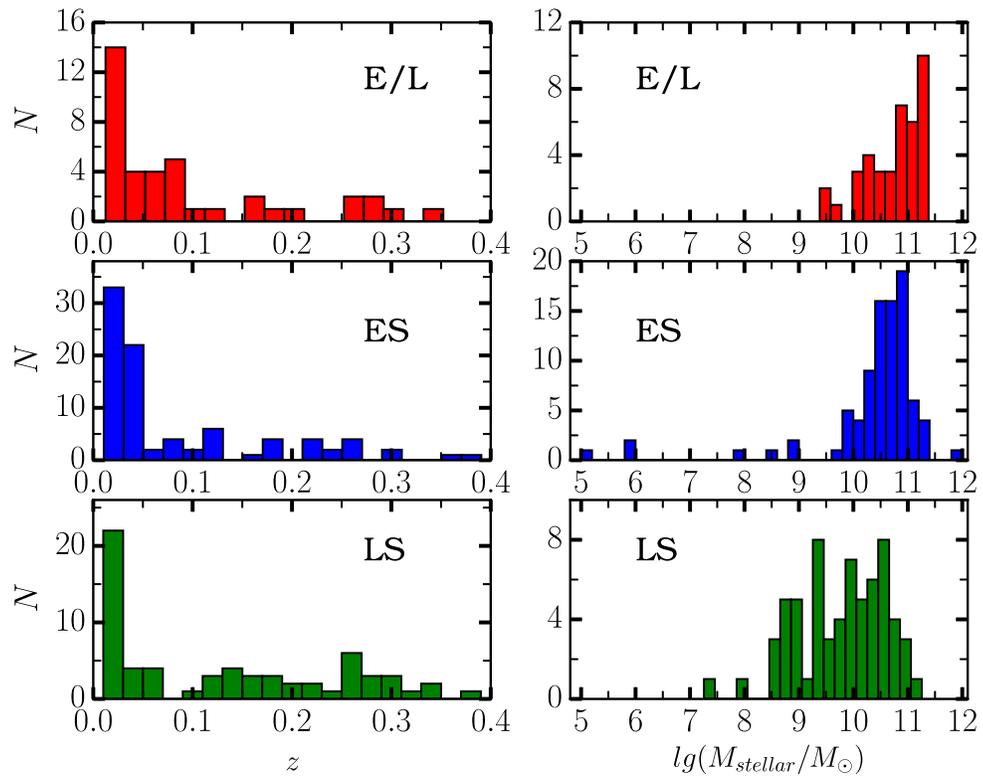}
\end{center}
   \caption{Redshift (left) and stellar mass (right) distributions for different host galaxy morphologies.}
  \label{fig:Host_mass}
\end{figure}

\begin{figure}
\begin{center}
  \includegraphics[width = 15cm]{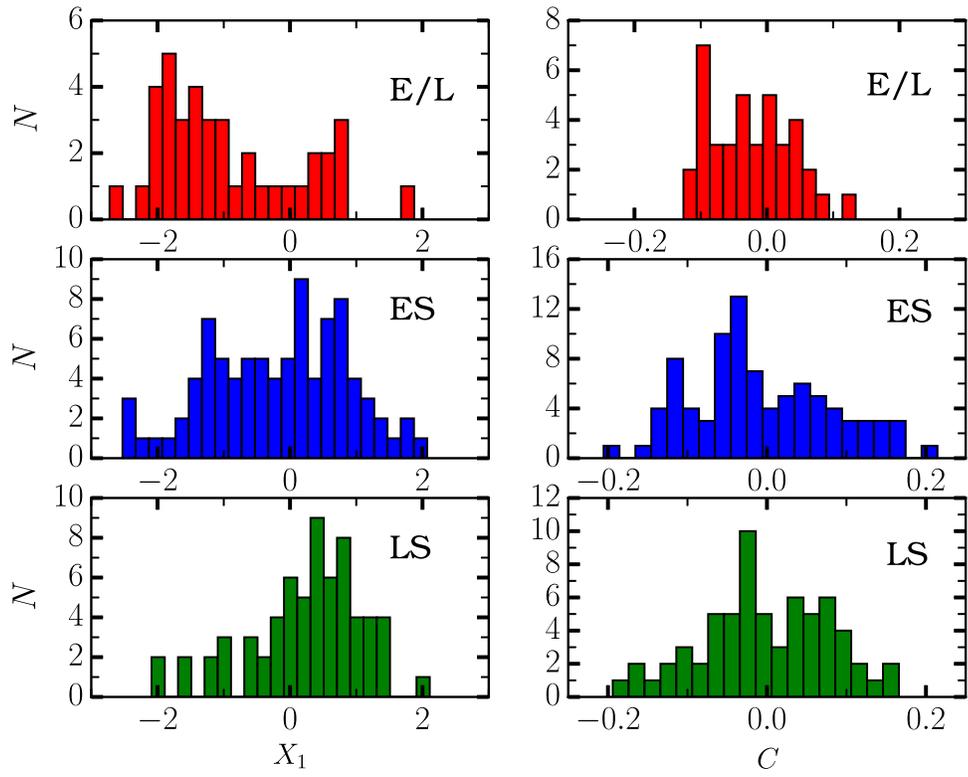}
\end{center}
   \caption{The distribution of the SALT2 parameters, stretch $X_1$ (left) and colour $C$ (right) for different host galaxy morphologies.}
  \label{fig:str_col}
\end{figure}

The results of this analysis are represented in Table~\ref{table:SALT2_Morpho} and visualized in Fig.~\ref{fig:str_col}. The trend in stretch behavior is found. Low-stretch supernovae appear in E/L galaxies, the intermediate-stretch supernovae are related to ES galaxies, and the highest average stretch belongs to supernovae in LS hosts. If we assume that star formation rate connects with host galaxy morphology, so that low-star formation rate is common for elliptical galaxies, and that high-star forming regions are usually located in arms of spiral galaxies, our results are consistent with~\cite{2006ApJ...648..868S,2009ApJ...707.1449N,2010MNRAS.406..782S,2012ApJ...755...61S,2013MNRAS.435.1680J}. They found that low-stretch supernovae are preferentially hosted by galaxies with little or no ongoing star formation. Moreover,~\cite{2010MNRAS.406..782S,2013MNRAS.435.1680J} showed that fast decline rates (low-stretch) SNe~Ia are almost absent in low-$M_{stellar}$ galaxies and usually happen in more massive galaxies. This result is consistent with ours, as illustrated by Fig.~\ref{fig:Host_mass} showing that E/L hosts have the highest stellar mass, on average.   

In addition, a small trend between SN colour and host galaxy type is identified. SNe~Ia in E/L galaxies have slightly bluer colour than others (see Table~\ref{table:SALT2_Morpho}). This could be explained by the fact that early-type and late-type spirals contain more dust which make the SN colour redder. However, the errors are too large to allow a conclusion. 

\subsection{Hubble residuals and RMS}
We apply two different approaches. First, we examine the residuals for each subsample of SNe~Ia from one global fit in which the $\alpha$, $\beta$, and $M_{\rm B}$ parameters are common for the entire sample. 
The second approach examines the residuals and the variation of the three nuisance parameters, $M_{\rm B}$, $\alpha$, and $\beta$, by performing a separate fit for each subsample (e.g.,~\cite{2010MNRAS.406..782S}). We aim to investigate if SNe~Ia could be physically different in the separate samples due to environmental effects. Therefore, we assume constant cosmological parameters in both approaches ($\Omega_\Lambda = 0.705$~\cite{Betoule2014}). 

\begin{figure}
\begin{center}
  \includegraphics[width = 15cm]{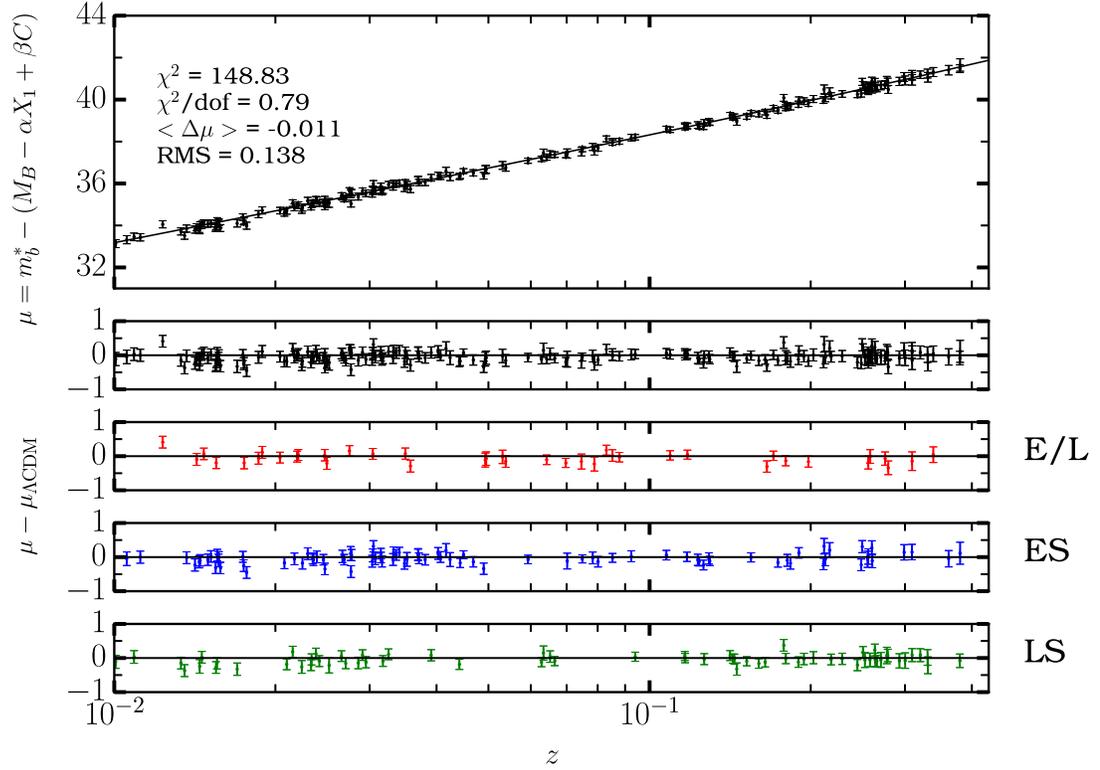}
\end{center}
   \caption{The Hubble diagram for the full selected sample of SN~Ia (192 SNe) with identified host galaxy morphology. The residuals for the three morphological subsamples of SN~Ia from one global fit in which the $\alpha$, $\beta$, and $M_{\rm B}$ parameters are common for the entire sample are presented. The cosmology is fixed at $\Omega_\Lambda = 0.705$.}
  \label{fig:HD}
\end{figure}

The Hubble diagram for the full selected sample of 192 SNe~Ia is presented in~Fig.~\ref{fig:HD}. The different subsamples are indicated by different colours. The average residuals $\langle \Delta \mu\rangle$ for all three types of host are compatible with zero (see Table~\ref{table:morphofree}). However, the RMS in E/L hosts is a bit higher than the one in LS galaxies. This result suggests that SN~Ia could be different in the considered types of galaxies. 

\begin{table}[h]
 \noindent\makebox[\textwidth]{
\begin{tabular}{lcccccc}
  \hline
  & $\langle \Delta \mu\rangle$ & $\Delta \mu$ RMS & $\chi^{2}$/dof &  $\alpha$ & $\beta$ & $M_{\rm B}$\\
 \hline 
  Our sample & $-0.011 \pm 0.010$ & $0.138 \pm 0.007$ & $0.79$ & $0.136 \pm 0.011$ & $2.74 \pm 0.14$ & $-19.09 \pm 0.02$\\
  \hline
 E/L& $-0.022\pm 0.023$ & $0.144 \pm 0.016$ & $0.92$ &  $0.136$ & $2.74$ & $-19.09$\\
 ES& $-0.004 \pm 0.016$ & $0.146 \pm 0.011$ & $0.87$ & $0.136$ & $2.74$ & $-19.09$\\
 LS& $-0.014 \pm 0.016$ & $0.125 \pm 0.011$ & $0.69$  & $0.136$ & $2.74$ & $-19.09$\\
 \hline
 E/L& $-0.001 \pm 0.023$ & $0.143 \pm 0.016$ & $0.86$  & $0.165\pm0.025$ & $2.75\pm0.39$ & $-19.14\pm0.04$\\
 ES& $-0.005 \pm 0.016$ & $0.147 \pm 0.011$ & $0.86$ & $0.139 \pm 0.016$ & $2.91 \pm 0.21$ & $-19.09\pm0.02$\\
 LS& $-0.011 \pm 0.014$ & $0.115 \pm 0.010$ & $0.64$ & $0.105\pm0.023$ & $2.50\pm0.24$ & $-19.10\pm0.03$\\
 \hline
\end{tabular}}
\caption{The results of Hubble diagram fitting for the full selected sample and two approaches we used, with fixed and free $\alpha$, $\beta$ and $M_{\rm B}$ parameters. $\langle \Delta \mu\rangle$ is the average of the residuals. The cosmology is fixed at $\Omega_\Lambda = 0.705$.}
  \label{table:morphofree}
\end{table}

\begin{figure}
\begin{center}
  \includegraphics[width = 15cm]{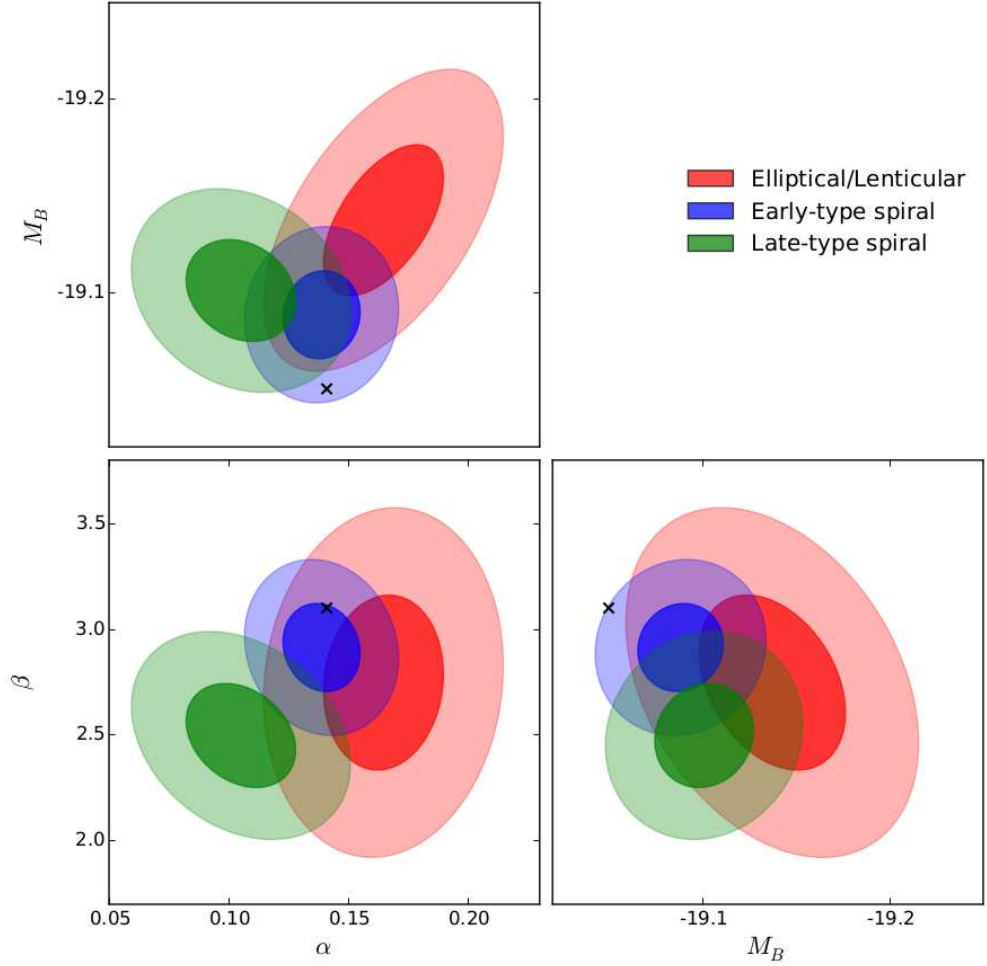}
\end{center}
   \caption{Joint confidence contours (1- and 2-$\sigma$) in two-parameter plots of $\alpha$, $\beta$ and $M_{\rm B}$ for fits where the SNe are split according to the host galaxy morphology. The black crosses represent the best fitting values ($\alpha = 0.141$, $\beta = 3.101$, $M_{\rm B} = -19.05$) from~\cite{Betoule2014}. The cosmology is fixed at $\Omega_\Lambda = 0.705$.}
   \label{fig:CP}
\end{figure}

To better understand the reason of the RMS variations we perform the Hubble diagram fitting for each subsample of supernovae separately. The fitting parameters are collected in Table~\ref{table:morphofree}. Fig.~\ref{fig:CP} shows  the joint confidence contours in the three combinations of $\alpha$, $\beta$ and $M_{\rm B}$ parameters for each SN subsamples. $M_{\rm B}$ parameter is almost left unchanged under the uncertainties. The shift in the $\beta$ parameter is also included in the uncertainties. On the other hand, there is a systematic shift of the value of the $\alpha$ parameter. Using the current values of the nuisance parameters and the average values of $X_1$ and $C$ for each subsample, we calculate the average absolute magnitudes in $B$-band: $\langle M_{\rm E/L}\rangle = -19.05 \pm 0.11$, $\langle M_{\rm ES}\rangle = -19.09 \pm 0.05$, $\langle M_{\rm LS}\rangle = -19.13 \pm 0.05$.
The obtained result does not agree with results from previous studies where after light-curve and colour corrections SN~Ia are brighter in E/S0 galaxies or in low star-formation hosts than those in late-type spirals or high star-formation hosts~\cite{2009ApJ...700.1097H,2010MNRAS.406..782S}. The $\alpha$, $\beta$, and $M_{\rm B}$ parameters parametrize the luminosity variations within the SN~Ia samples and are likely related to the physics of the SN~Ia explosion and/or the SN~Ia environment. Therefore, the found difference even in one parameter ($\alpha$) emphasizes the importance of the environment in SN~Ia studies. 

SNe~Ia from LS galaxies (star-forming) appear to be more homogeneous than the others. The Hubble residual dispersion is smaller for them ($0.115\pm0.010$) in comparison with the full sample residual dispersion ($0.138\pm0.007$). This result suggests that a division of SNe~Ia by types according to the morphology of their host galaxies would improve the future cosmological analysis. Previous studies supported the idea that SNe~Ia in locally star-forming regions are more appropriate for cosmology due to their low brightness dispersion~\cite{Rigault2013}. 

\section{Conclusions}
We derived the host galaxy morphology for 192 SNe~Ia from the nearby and SDSS JLA sample. The supernovae were divided into three subsamples depending on host morphology:  elliptical/lenticular (E/L), early-type spiral (ES), and late-type spiral galaxies (LS). 

Our main conclusions are:

1) As was previously shown by other studies, the SN~Ia stretch parameter is closely correlated with host morphology. The low stretch (fast decline) SNe usually exploded in E/L galaxies, i.e. in old stellar population environment.

2) We did not find a statistically significant trend for the colour parameter. 

3) We found a trend in the stretch nuisance parameter $\alpha$. Its value decreases from E/L to LS galaxies. 

4) We saw that, in an old stellar population and with low dust environment, supernovae are intrinsically fainter after stretch and colour corrections. This conclusion contradicts the results obtained by~\cite{2009ApJ...700.1097H} (see also~\cite{2010MNRAS.406..782S} and references in it). However, our results are consistent within 1-$\sigma$ for all subsamples. This can be improved by adding SNLS JLA supernovae in the analysis.

5) The host galaxy morphology affects the residual dispersion in the Hubble diagram. We noticed that SNe~Ia in LS are more homogeneous with RMS~$ = 0.115\pm0.010$. 

\bigskip
The variation of the light curve parameters, especially the luminosity, with morphological type of host galaxy influences the cosmological analysis (e.g.,~\cite{2009ApJ...700.1097H,Rigault2013}). We conclude that for further cosmological studies the environmental effects should be taken into account.

\section{Acknowledgements} 
The authors thank Marc Betoule for the discussion and technical issue, and acknowledge Rick Kessler, Masao Sako, and Ravi Gupta for their help with SDSS data. We thank the reviewer for his/her thorough review and highly appreciate the comments and suggestions. The authors acknowledge the Universit\'e Blaise Pascal, the French CNRS-IN2P3 agency and the R\'egion Auvergne-Rh\^one-Alpes for their funding support. The work of M.V.P. (writing of the paper and comparison of the results with previously published results) was supported by RSCF grant no. 14-12-00146. 
\bibliography{references}
\bibliographystyle{ieeetr}

\newpage
\section*{Appendix}
\begin{center}
 \begin{longtable}{lccc}
\caption{Final sample of SN~Ia used in current analysis. Host types are divided by three categories: elliptical/lenticular (E/L), early-type spiral (ES), and late-type spiral (LS). For SDSS galaxies the name is IDobj taken from SDSS Data Release 8~\cite{2011ApJS..193...29A}.} \label{SN_list} \\
 \hline
 SN name & Host name & $z_{\rm cmb}$ & Host type\\
 \hline
sn1999ac	&	NGC6063	&	0.01006	&	LS	\\

sn1997do	&	UGC3845	&	0.01055	&	ES	\\

sn2002dp	&	NGC7678	&	0.010888	&	LS	\\

sn2006bh	&	NGC7329	&	0.011184	&	ES	\\

sn2005al	&	NGC5304	&	0.012317	&	E/L	\\

sn2001ep	&	NGC1699	&	0.013324	&	LS	\\

sn1999dq	&	NGC976	&	0.013533	&	LS	\\

sn2002ha	&	NGC6962	&	0.013652	&	ES	\\

sn2001bt	&	IC4830	&	0.014151	&	ES	\\

sn1997e	&	NGC2258	&	0.014256	&	E/L	\\

sn1992al	&	ESO234-69	&	0.014411	&	ES	\\

sn1999dk	&	UGC1087	&	0.01444	&	LS	\\

sn2005bo	&	NGC4708	&	0.014467	&	ES	\\

sn2001fe	&	UGC5129	&	0.014575	&	LS	\\

sn2006n	&	MCG+11-08-012	&	0.0147	&	E/L	\\

sn2005kc	&	NGC7311	&	0.01496	&	ES	\\

sn2005el	&	NGC1819	&	0.015154	&	ES	\\

sn2004eo	&	NGC6928	&	0.015469	&	ES	\\

sn2001cn	&	IC4758	&	0.015487	&	LS	\\

sn2001bf	&	2MASXJ18013361+2615109	&	0.015504	&	E/L	\\

sn2001cz	&	NGC4679	&	0.015554	&	LS	\\

sn1999aa	&	NGC2595	&	0.015573	&	ES	\\

sn1994s	&	NGC4495	&	0.015664	&	ES	\\

sn2004ey	&	UGC11816	&	0.015678	&	ES	\\

sn2001v	&	NGC3987	&	0.015791	&	ES	\\

sn1996bv	&	UGC3432	&	0.016966	&	LS	\\

sn2006ax	&	NGC3663	&	0.017387	&	ES	\\

sn1998v	&	NGC6627	&	0.01745	&	ES	\\

sn2000dk	&	NGC382	&	0.017483	&	E/L	\\

sn1998ef	&	UGC646	&	0.017672	&	ES	\\

sn2007ci	&	NGC3873	&	0.018599	&	E/L	\\

sn1992bo	&	ESO352-57	&	0.018902	&	E/L	\\

sn2005ki	&	NGC3332	&	0.020396	&	E/L	\\

sn1992bc	&	ESO300-9	&	0.020792	&	ES	\\

sn2003w	&	UGC5234	&	0.020997	&	LS	\\

sn2002jy	&	NGC477	&	0.021541	&	LS	\\

sn2007bc	&	UGC6332	&	0.021706	&	ES	\\

sn2006bq	&	NGC6685	&	0.021956	&	E/L	\\

sn2008bf	&	NGC4055	&	0.022068	&	E/L	\\

sn1995ak	&	IC1844	&	0.022399	&	LS	\\

sn2006et	&	NGC232	&	0.022466	&	ES	\\

sn2006ar	&	MCG+11-13-036	&	0.02298	&	ES	\\

sn2006cp	&	UGC7357	&	0.023321	&	LS	\\

sn2000ca	&	2MASXJ13352270-3409338	&	0.02336	&	LS	\\

sn2006ac	&	NGC4619	&	0.023471	&	ES	\\

sn2000cn	&	UGC11064	&	0.023552	&	ES	\\

sn2007f	&	UGC8162	&	0.02381	&	LS	\\

sn1993h	&	ESO445-066	&	0.023909	&	ES	\\

sn2006sr	&	UGC14	&	0.024159	&	LS	\\

sn2005bg	&	MCG+03-31-093	&	0.024586	&	ES	\\

sn2002he	&	UGC4322	&	0.024725	&	E/L	\\

sn2002bf	&	2MASXJ10154226+5540030	&	0.024746	&	ES	\\

sn1994m	&	NGC4493	&	0.024951	&	E/L	\\

sn1992ag	&	ESO508-67	&	0.025198	&	LS	\\

sn2005ms	&	UGC04614	&	0.026567	&	LS	\\

sn1999gp	&	UGC01993	&	0.026677	&	ES	\\

sn2005na	&	UGC3634	&	0.026881	&	ES	\\

sn2007co	&	MCG+05-43-016	&	0.027064	&	LS	\\

sn2004gs	&	MCG+03-22-020	&	0.027499	&	E/L	\\

sn1992p	&	IC3690	&	0.027637	&	ES	\\

sn1998ab	&	NGC4704	&	0.027688	&	ES	\\

sn2002de	&	NGC6104	&	0.027748	&	ES	\\

sn2003u	&	NGC6365A	&	0.028561	&	LS	\\

sn2005eq	&	MCG-01-09-006	&	0.029086	&	LS	\\

sn2006qo	&	UGC4133	&	0.029526	&	LS	\\

sn1990o	&	MCG+03-44-003	&	0.030325	&	ES	\\

sn2001ba	&	2MASXJ11380048-3219286	&	0.030381	&	ES	\\

sn2003ch	&	2MASXJ07175757+0941218	&	0.030416	&	E/L	\\

sn1997dg	&	SDSSJ234014.21+261211.8	&	0.030482	&	ES	\\

sn1996c	&	MCG+08-25-047	&	0.030721	&	ES	\\

sn2006bt	&	MCG+03-41-004	&	0.031484	&	ES	\\

sn2006az	&	NGC4172	&	0.031485	&	ES	\\

sn2007bd	&	UGC4455	&	0.031638	&	ES	\\

sn1999cc	&	NGC6038	&	0.031764	&	LS	\\

sn2004as	&	SDSSJ112538.81+224952.2	&	0.032538	&	LS	\\

sn2004l	&	MCG+03-27-038	&	0.033173	&	ES	\\

sn2006s	&	UGC7934	&	0.033255	&	ES	\\

sn2006gr	&	UGC12071	&	0.033614	&	ES	\\

sn2005iq	&	MCG-03-01-008	&	0.034074	&	ES	\\

sn1996bl	&	2MASXJ00361813+1123354	&	0.034854	&	ES	\\

sn2003iv	&	2MASXJ02500884+1250372	&	0.034971	&	E/L	\\

sn1992bg	&	2MASXJ07415700-6231144	&	0.035218	&	ES	\\

sn2002hd	&	2MASXJ08540470-0710597	&	0.035755	&	E/L	\\

sn2000cf	&	2MASXJ15525577+6556079	&	0.036934	&	ES	\\

sn2006mo	&	MCG+06-02-017	&	0.036969	&	ES	\\

sn2001eh	&	UGC1162	&	0.037075	&	ES	\\

sn2002hu	&	2MASXJ02181834+3727513	&	0.038086	&	ES	\\

sn1999aw	&	SCPJ110136.37-060631.6	&	0.039083	&	LS	\\

sn2001az	&	UGC10483	&	0.040259	&	ES	\\

sn2005lz	&	UGC1666	&	0.040689	&	ES	\\

sn1992bh	&	LSBG-F119-024	&	0.041657	&	ES	\\

sn1992bl	&	ESO291-11	&	0.0424	&	ES	\\

sn2005hc	&	MCG+00-06-003	&	0.04492	&	ES	\\

sn2004gu	&	2MASXJ12462478+1156577	&	0.046856	&	ES	\\

sn1995ac	&	2MASXJ22453410-0845026	&	0.048984	&	ES	\\

sn1993ag	&	2MASXJ10033546-3527410	&	0.049325	&	E/L	\\

sn1990af	&	2MASXJ21345926-6244143	&	0.049519	&	E/L	\\

sn1993o	&	2MASXJ13310895-3312576	&	0.053126	&	E/L	\\

sn1998dx	&	UGC11149	&	0.05386	&	E/L	\\

sn2006ob	&	UGC1333	&	0.059289	&	ES	\\

sn1992bs	&	2MASXJ03292797-3716222	&	0.063345	&	LS	\\

sn2006an	&	SDSSJ121438.73+121347.7	&	0.065193	&	LS	\\

sn2006on	&	SDSSJ215558.49-010413.0	&	0.069751	&	E/L	\\

sn1993b	&	PGC662260	&	0.070153	&	ES	\\

sn1992ae	&	[WM92]212426.8-614612	&	0.074579	&	E/L	\\

sn2005ir	&	PGC3116670	&	0.07487	&	ES	\\

sn1999bp	&	PGC998264	&	0.078317	&	ES	\\

sn1992bp	&	2MASXJ03363839-1821123	&	0.07884	&	E/L	\\

sn2005ag	&	2MASXJ14564322+0919361	&	0.080103	&	ES	\\

SDSS10805	&	1237666407898940000	&	0.04413	&	LS	\\

SDSS14279	&	1237666339726690000	&	0.04428	&	ES	\\

SDSS3901	&	1237663784203910000	&	0.062818	&	LS	\\

SDSS10028	&	1237666339726360000	&	0.064229	&	E/L	\\

SDSS6057	&	1237666299482800000	&	0.06651	&	LS	\\

SDSS12781	&	1237657189833840000	&	0.08304	&	E/L	\\

SDSS722	&	1237657191979290000	&	0.08522	&	E/L	\\

SDSS3592	&	1237663204922490000	&	0.08528	&	ES	\\

SDSS21502	&	1237656906352300000	&	0.08784	&	E/L	\\

SDSS774	&	1237657069547680000	&	0.09243	&	ES	\\

SDSS2102	&	1237656568648830000	&	0.09401	&	LS	\\

SDSS21034	&	1237678617430130000	&	0.10751	&	ES	\\

SDSS7147	&	1237657190900830000	&	0.10924	&	E/L	\\

SDSS5395	&	1237663784219120000	&	0.11635	&	LS	\\

SDSS8719	&	1237663783127020000	&	0.11651	&	LS	\\

SDSS2561	&	1237678437019290000	&	0.11741	&	ES	\\

SDSS19953	&	1237663479795480000	&	0.11767	&	E/L	\\

SDSS2916	&	1237678617398540000	&	0.12303	&	ES	\\

SDSS6406	&	1237660024523920000	&	0.12376	&	ES	\\

SDSS2992	&	1237663237667550000	&	0.12608	&	ES	\\

SDSS744	&	1237663543682530000	&	0.12649	&	LS	\\

SDSS1032	&	1237666302164660000	&	0.12903	&	ES	\\

SDSS5751	&	1237663204919280000	&	0.12953	&	ES	\\

SDSS1794	&	1237663542603810000	&	0.14139	&	LS	\\

SDSS2635	&	1237663237129500000	&	0.1431	&	LS	\\

SDSS8921	&	1237663543143830000	&	0.14401	&	LS	\\

SDSS11300	&	1237657190371300000	&	0.14547	&	LS	\\

SDSS2031	&	1237656567038150000	&	0.15186	&	LS	\\

SDSS5550	&	1237657191443660000	&	0.15473	&	ES	\\

SDSS3317	&	1237657071695820000	&	0.15999	&	LS	\\

SDSS3087	&	1237666338116930000	&	0.1644	&	LS	\\

SDSS12843	&	1237657189815680000	&	0.16565	&	E/L	\\

SDSS12856	&	1237663544220980000	&	0.17028	&	E/L	\\

SDSS3080	&	1237666338115360000	&	0.17374	&	ES	\\

SDSS5635	&	1237663543147360000	&	0.1781	&	LS	\\

SDSS2372	&	1237657070091110000	&	0.17962	&	E/L	\\

SDSS6936	&	1237656567579870000	&	0.17969	&	LS	\\

SDSS17215	&	1237666302168200000	&	0.18079	&	ES	\\

SDSS8213	&	1237656906354000000	&	0.18327	&	ES	\\

SDSS6304	&	1237678617429410000	&	0.18965	&	LS	\\

SDSS762	&	1237666338114770000	&	0.1901	&	ES	\\

SDSS2246	&	1237666299481750000	&	0.19422	&	LS	\\

SDSS15222	&	1237657191980200000	&	0.19801	&	E/L	\\

SDSS7243	&	1237663479793390000	&	0.20231	&	LS	\\

SDSS7847	&	1237666407919780000	&	0.21134	&	ES	\\

SDSS2330	&	1237678434328180000	&	0.21179	&	ES	\\

SDSS8495	&	1237656567585180000	&	0.21295	&	ES	\\

SDSS9467	&	1237678595929410000	&	0.21698	&	ES	\\

SDSS5533	&	1237663543682270000	&	0.21829	&	LS	\\

SDSS3452	&	1237663544221760000	&	0.22893	&	LS	\\

SDSS3377	&	1237666302167880000	&	0.24448	&	LS	\\

SDSS3451	&	1237663544221500000	&	0.24851	&	ES	\\

SDSS3199	&		&	0.24961	&	ES	\\

SDSS5717	&	1237666339189560000	&	0.25037	&	LS	\\

SDSS9032	&	1237663542612590000	&	0.25249	&	LS	\\

SDSS9457	&	1237663543685420000	&	0.25539	&	ES	\\

SDSS1112	&	1237663478724430000	&	0.25611	&	ES	\\

SDSS17340	&	1237666408460450000	&	0.25617	&	E/L	\\

SDSS6108	&	1237663543696750000	&	0.258	&	LS	\\

SDSS8046	&	1237663784751400000	&	0.25833	&	E/L	\\

SDSS2017	&	1237663479793710000	&	0.26014	&	ES	\\

SDSS1253	&	1237663457779380000	&	0.26059	&	ES	\\

SDSS2422	&	1237657191979810000	&	0.26349	&	LS	\\

SDSS2943	&		&	0.26405	&	LS	\\

SDSS6315	&	1237659743703800000	&	0.26576	&	LS	\\

SDSS6192	&	1237678617412760000	&	0.27044	&	LS	\\

SDSS4046	&	1237657191976530000	&	0.27544	&	E/L	\\

SDSS4000	&	1237663783674120000	&	0.27745	&	LS	\\

SDSS5957	&	1237663783675760000	&	0.27852	&	LS	\\

SDSS6196	&	1237663542612460000	&	0.27916	&	E/L	\\

SDSS6249	&	1237657190369790000	&	0.29287	&	LS	\\

SDSS6137	&		&	0.29888	&	ES	\\

SDSS5391	&	1237663237129310000	&	0.30021	&	LS	\\

SDSS6699	&	1237656567042800000	&	0.30915	&	ES	\\

SDSS5844	&		&	0.30929	&	LS	\\

SDSS16211	&	1237666408437250000	&	0.30933	&	E/L	\\

SDSS7475	&		&	0.32045	&	LS	\\

SDSS4241	&	1237657189836850000	&	0.33051	&	LS	\\

SDSS4679	&	1237663204923610000	&	0.33103	&	LS	\\

SDSS2533	&	1237663783674180000	&	0.3388	&	E/L	\\

SDSS4577	&	1237666408459270000	&	0.36193	&	ES	\\
SDSS7779	&	1237663543137270000	&	0.37986	&	LS	\\
SDSS11206	&	1237657192517270000	&	0.38026	&	ES	\\
\hline
\end{longtable}
\end{center}

\end{document}